\documentclass[useAMS]{mn2e}
\usepackage{times}
\input{epsf}

\title[Spectral transitions from ASM]
{X-ray spectral transitions of black holes from {\it RXTE\/} All-Sky
Monitor}

\author[M. Gierli{\'n}ski and J. Newton]
{Marek
Gierli{\'n}ski$^{1,2}\thanks{E-mail:Marek.Gierlinski@durham.ac.uk}$ and
Jo Newton$^{1}$\\
$^1$Department of Physics, University of Durham, South Road,
Durham DH1 3LE, UK\\
$^2$Astronomical Observatory, Jagiellonian University, Orla 171,
30-244 Krak{\'o}w, Poland\\
}

\date{Submitted to MNRAS}
\pagerange{\pageref{firstpage}--\pageref{lastpage}} \pubyear{2005}

\begin{document}

\topmargin = -0.5cm

\maketitle

\label{firstpage}

\begin{abstract}

We have analysed X-ray outbursts from several Galactic black hole (GBH)
transients, as seen by the All-Sky Monitor (ASM) on board {\it Rossi
X-ray Timing Explorer} ({\it RXTE}). We have used the best estimates of
distance and black hole mass to find their luminosity (scaled to the
Eddington limit), which allowed for direct comparison of many sources.
We have found two distinct hard-to-soft state transitions in the
initial part of the outburst. The distinction is made on the basis of
the transition luminosity, its duration, the shape of the track in the
hardness-luminosity diagram, and evolution of the hardness ratio. The
bright/slow transition occurs at $\sim$30 per cent of Eddington
(estimated bolometric) luminosity and takes $\ga$30 days, during which
the source quickly reaches the intermediate/very high state and then
proceeds to the soft state at much slower pace. The dark/slow
transition is less luminous ($\la$10 per cent of Eddington), shorter
($\la$15 days) and the source does not slow its transition rate before
reaching the soft state. We speculate that the distinction is due to
irradiation and evaporation of the disc, which sustains the
Comptonizing corona in the bright intemediate/very high state.

\end{abstract}

\begin{keywords}

accretion, accretion discs -- instabilities -- X-rays: binaries

\end{keywords}

\section{Introduction}
\label{sec:introduction}

Black holes are remarkably simple objects, entirely characterized by
their mass and spin, so we should expect similar appearance of their
accretion flows. In Galactic binaries black holes have comparable
masses, so the accretion flow properties should be then determined by
their spin and (instantaneous) accretion rate. It has been established
recently that the accretion rate history plays an important role in
shaping the accretion flow as well (e.g. van der Klis 2001; Maccarone
\& Coppi 2003). The inclination angle can also have some, but rather
weak, effect on the observed properties, due to anisotropy of the disc
or corona emission. Most of these effects can be easily accounted for:
the mass by scaling the observed luminosity to Eddington ($L_{\rm
Edd}$) limit, the accretion history by tracing the entire outburst of a
transient, the inclination angle effects are easy to understand within
existing models of accretion. Perhaps the most challenging are the
effects of black hole spin as they involve the properties of the
accretion flow near and below the last stable orbit (e.g. Done \&
Gierli{\'n}ski 2006).

GBHs are observed in several distinct X-ray spectral states (see
Zdziarski \& Gierli{\'n}ski 2004; McClintock \& Remillard in press and
references therein). This is best seen in X-ray transients undergoing
occasional outbursts, when their accretion rate (and luminosity) can
change by several orders of magnitude. With varying accretion rate the
X-ray spectrum can also change dramatically, passing through various
states on timescales from days to months. These spectral transitions
are probably the result of changes in the geometry of the accretion
flow (for a review see Done \& Gierli{\'n}ski 2004).

Typically, a transient begins its outburst in the low/hard (LH) state,
where its spectrum can be roughly described by a hard (photon spectral
index $\Gamma \sim$ 1.5--2) power law with high-energy cutoff at $\sim$
100 keV (e.g. Wilson \& Done 2001). The LH state can be seen at
luminosities up to $\sim$0.2 $L_{\rm Edd}$ during the rise of an
outburst (Done \& Gierli{\'n}ski 2003), but is less bright ($\la$0.04
$L/L_{\rm Edd}$) in the decaying part of the outburst (Maccarone 2003).
A truncated disc, replaced by the optically thin, hot, Comptonizing
inner flow can explain the hard state (e.g. Poutanen, Krolik \& Ryde
1997; Esin et al. 2001), but other geometries have been also proposed
(e.g. Young et al. 2001; Beloborodov 1999).

After the initial LH state the X-ray spectrum usually undergoes
transition into the high/soft (HS) state, via intermediate (IM) or very
high (VH) state. The X-ray spectrum is then dominated by the disc
emission of temperature of $\sim$ 1 keV, accompanied by a soft ($\Gamma
\sim$ 2--2.5) power-law tail to higher energies. In the soft state the
cold disc probably extends down to or near the last stable orbit and
the hard component is produced by Comptonization in active regions or
corona above the disc surface (e.g. Gierli{\'n}ski et al. 1999).

The IM/VH state is characterized by a soft X-ray spectrum dominated by
a steep ($\Gamma \sim$ 2.5) power law, extending to several hundred keV
without apparent cutoff (Grove et al. 1998). Initially it was
recognized as the brightest state of a given object (Miyamoto et al.
1991), but HS states brighter then IM/VH state were seen (e.g.
Zdziarski et al. 2001). IM/VH state has been observed in various
sources at a wide range of luminosities of $\sim$0.01--1 $L_{\rm Edd}$.
Another characteristic feature of IM/VH state is strong rapid aperiodic
variability often accompanied by a distinct quasi-periodic oscillation.

The IM/VH state is typically related to the transitions between the
hard and soft states (Rutledge et al. 1999). In many transients it has
been seen after the initial hard state, when the source reached high
luminosity ($\ga$0.1 $L_{\rm Edd}$). This intermediate phase can last
up to several dozen days before the source moves to the standard HS
state. There are, however, instances of IM/VH state observed during
`failed' transitions or between two periods of the soft state (e.g.
Homan et al. 2001; Kubota \& Done 2004).

The X-ray spectral states and transitions of Galactic black holes were
analysed from the {\it RXTE} spectral data by many authors (e.g.
Sobczak et al. 2000; Homan et al. 2001; Done \& Gierli{\'n}ski 2003;
Fender, Belloni \& Gallo 2004; Homan \& Belloni 2005). A lot of work
has been also done on the {\it RXTE} ASM data (e.g. Grimm, Gilfanov \&
Sunyaev 2002; Zdziarski et al. 2002; Maccarone \& Coppi 2003; {\v
S}imon 2004; Zdziarski et al. 2004; McClintock \& Remillard in press).
However, few of these works attempted to discuss properties of spectral
states in terms of absolute {\em luminosity}. In this paper we present
the first systematic study of black hole X-ray spectral states, based
on hardness and luminosity derived from {\it RXTE} ASM data. We use the
best known estimates of distance and mass of accreting black holes to
derive their luminosity as a fraction of $L_{\rm Edd}$. This allows us
for direct comparison between different sources. We also trace {\em
spectral evolution\/} during the outburst. Using these data we show
that there are two distinct types of transition between the hard and
soft spectral states in the initial part of the outburst.

\section{Data reduction}
\label{sec:data}

\begin{table*}
\begin{tabular}{lcc}
\hline Source Name & $M$ (M$_\odot$) & $D$ (kpc)\\
\hline
4U 1543--47       & 9.4 (7.4--11.4)$^a$  & 7.5 (7--8)$^a$ \\
XTE J1550--564    & 10 (9.7--11.6)$^b$   & 5.3 (2.8--7.6)$^b$ \\
XTE J1650--500    & 4 ($\la$7.3)$^c$    & 2.6 (1.9--3.3)$^d$ \\
GX 339--4         & 6 (2.5--10)$^e$      & 8 (6.7--9.4)$^f$ \\
XTE J1739--278    & [10]             & 8.5 (6--11)$^g$ \\
H 1743--322       & [10]             & [8.5] \\
XTE J1859+226     & [10]             & 7.6 (4.6--8)$^h$ \\
XTE J2012+381     & [10]             & [8.5]\\
\hline
\end{tabular}

\caption{The list of the sources used in this paper, together with
assumed mass ($M$) and distance ($D$) estimates and their
uncertainties. The numbers in square brackets refer to quantities that
were not known in the time of writing of this paper and were assumed:
mass of 10 M$_\odot$ and distance of 8.5 kpc. The numbered references
are as follows: [a] Park et al. (2004) [b] Orosz et al. (2002) [c]
Orosz et al. (2004) [d] Homan et al. (2006) [e] Cowley et al. (2002)
[f] Zdziarski et al. (2004) [g] Greiner, Dennerl \& Predehl (1996) [h]
Hynes et al. (2002).}

\label{tab:sources}
\end{table*}

We have analysed publicly available light curves of 8 Galactic black
hole candidates from ASM (Levine et al. 1996) on board {\it RXTE}
(Jahoda et al. 1996). These sources were selected because their data
had statistics sufficient to build hardness-luminosity diagrams. We
used one-day averages in three ASM energy bands, roughly corresponding
to 1.5--3 keV, 3--5 keV and 5--12 keV. To allow for direct comparison
between different sources, we converted the observed count rates into
luminosities.

First, we calculated the energy flux in each ASM channel, adopting the
method of Zdziarski et al. (2002), who built the ASM response matrix,
comparing various pointed Cyg X-1 observations with ASM count rates. To
check the validity of this method for the IM/VH spectral shape, which
is the main topic of this paper, we have used the physical model fitted
to the very high state {\it RXTE\/} spectrum of XTE J1550--564 (model
HYB in table 4 in Gierli{\'n}ski \& Done 2003). This model gave the
1.5--12 keV flux of $4.3\times10^{-8}$ erg s$^{-1}$ cm$^{-2}$, while
the corresponding ASM data converted into flux resulted in
$4.5\times10^{-8}$ erg s$^{-1}$ cm$^{-2}$, in excellent agreement with
the pointed observation.

Then, we found the total 1.5--12 keV (isotropic) luminosity and
expressed it as a fraction of Eddington luminosity [$L_{\rm Edd} =
1.26\times10^{38}$ ($M$/M$_\odot$) erg s$^{-1}$] using the best-known
estimates of mass and distance, summarized in Table \ref{tab:sources}.
We also calculated the hard flux (not count rate) hardness ratio, HR,
of energy bands 5--12/3--5 keV. For clarity, we have discarded the data
points with poor statistics (relative error on HR greater then 0.3)
from hardness-luminosity diagrams.

\section{Results}
\label{sec:results}

\begin{figure*}
\begin{center}
\leavevmode \epsfxsize=17cm \epsfbox{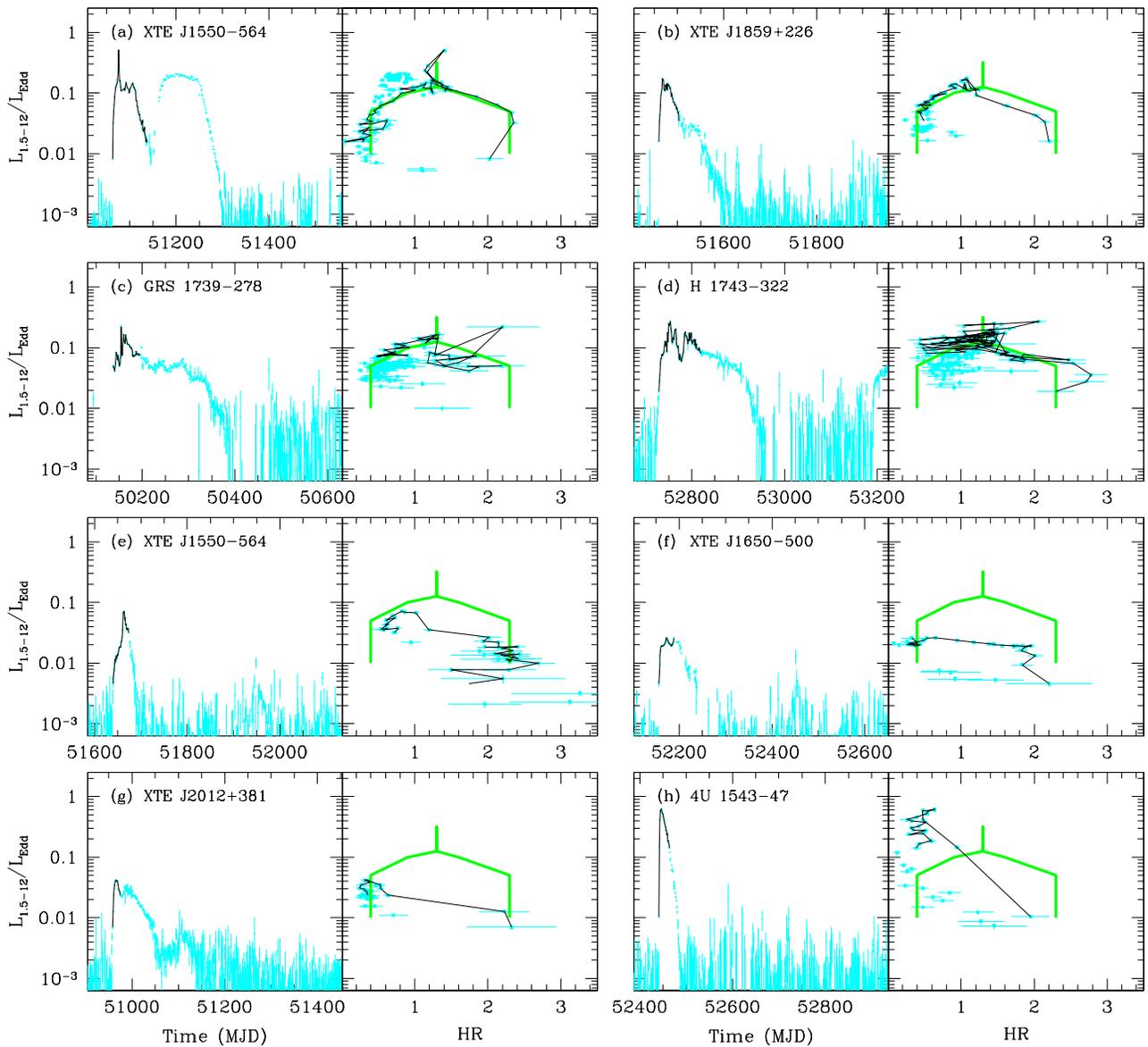}
\end{center}
\caption{Light curves and hardness-luminosity diagrams of the outbursts
of sources listed in Table \ref{tab:sources}, except for GX 339--4,
which is shown in Fig. \ref{gx}. The light curves are shown by light
grey (cyan in colour) crosses. The thin black curve follows the
transition from the hard to the soft state and joins the same data
points in the light curve and H-L plot. The thick grey (green in
colour) inverted y-shaped line in the H-L diagrams illustrates the
hard-to-soft transition of XTE J1550--564 and is used for comparison
with other sources.} \label{diags}
\end{figure*}

For our analysis we have selected ten outbursts of the transients
listed in Table \ref{tab:sources}. In Figs. \ref{diags} and \ref{gx} we
present their light curves and hardness-luminosity (H-L) diagrams. In
order to guide the eye and make comparison of various sources easier,
we have copied a schematic shape of the H-L track of XTE J1550--564
1998 outbursts (shown in Fig. \ref{diags}a) in all other H-L diagrams
(the inverted Y-shaped curve).

Fig. \ref{trans} shows the evolution of the hardness ratio of these
outbursts. The horizontal lines show positions of hard and soft
branches. These lines were used to estimate the duration of the
hard-to-soft transition, which was defined as the time between the last
point on the hard branch and the first point on the soft branch.
Certainly, this is -- to some extent -- arbitrary, since we have
defined these branches by reading the diagrams. The soft branch (and
hard branch, where possible) corresponds to hardness ratio which
stabilized at a certain value. Where only one or a few points were
available in the hard state, we used them as the hard-state HR. The
position of each branch varies slightly from source to source, but in
most cases they are easy to identify from the diagrams. We will discuss
possible drawbacks of this method in Section \ref{sec:caveats}.

We have estimated the transition luminosity as $L/L_{\rm Edd}$ at
hardness of HR = 1.25. We also found the fluence of each outburst,
defined as the dimensionless $L/L_{\rm Edd}$ integrated over the entire
outburst (so it has units of time). We summarize these results in Table
\ref{tab:transitions}.

These results suggest existence of two different types of hard-to-soft
state transition. The bright/slow (BS) transition [Figs.
\ref{diags}(a--d) and \ref{trans}(a--d)] is characterized by the
transition luminosity of $\sim$0.1 $L/L_{\rm Edd}$ and duration of
$\ga$30 days. In the H-L diagram they trace a semi-circular,
anti-clockwise track between the LH and HS branches. The best examples
of this transition are shown by XTE J1550--564 (1998 outburst) and XTE
J1859+226. The dark/fast (DF) transition [Figs. \ref{diags}(e--g) and
\ref{trans}(e--g)] is significantly dimmer ($\la$0.05 $L/L_{\rm Edd}$)
and faster ($\la$15 days). In the H-L diagram they move on an almost
straight line between the LH and HS branches. The best examples of this
transition are shown by XTE J1650-500 and XTE J2012+381. The last
column in Table \ref{tab:transitions} shows the derived type of the
transition. We discuss the details of this categorization in Sections
\ref{sec:BS} and \ref{sec:DF}.

\subsection{Bright/slow transition}
\label{sec:BS}

Fig.~\ref{diags}(a) shows the light curve and hardness-luminosity
diagram of the 1998 outburst of XTE J1550--564. Fig. \ref{trans}(a)
shows the evolution of its hardness ratio. This particular outburst had
a very good {\it RXTE} coverage of pointed observations and its
broad-band energy spectra were analysed in details (e.g. Sobczak et al.
2000; Gierli{\'n}ski \& Done 2003; Kubota \& Done 2004). It is known to
cover all spectral states (e.g. Homan et al. 2001). In Figs.
\ref{diags}(a) and \ref{trans}(a) we connect the data points of the
initial part (first 76 days) of the outburst. The outburst began in the
LH state (Wilson \& Done 2001) on the right-hand branch in the H-L
diagram (at HR $\sim$ 2.5). Then, as the luminosity increased, the
spectrum softened fairly quickly and the source moved anticlockwise in
the H-L diagram, reaching the top point corresponding to a bright flare
in the light curve. This region in the H-L diagram corresponds to the
IM/VH state (Gierli{\'n}ski \& Done 2003; Kubota \& Done 2004). Then,
the luminosity decreased at much slower rate, and spectrum softened
even more, while the source reached the HS (left-hand) branch in the
H-L diagram at HR $\sim$ 0.5. The entire transition from leaving the LH
branch until reaching the HS branch took 54 days.

Panels (a--d) of Figs. \ref{diags} and \ref{trans} show the results
from outbursts which we have categorized as the BS transition. They all
begin in the LH branch of the H-L diagram, travel to the top `cusp' at
similar luminosity of $\sim$0.1 $L_{\rm Edd}$ and then move on to the
HS branch. XTE J1550--564 and XTE J1859+226 had a flare (brighter in
the former source) around the `cusp'. In terms of the HR evolution
(Fig. \ref{trans}) they seem to begin the outburst with a fairly fast
transition to the IM/VH state at HR $\sim$1.3, after which the
transition to the soft branch is significantly slower. This is
illustrated in Fig. \ref{curves}. The transition to the IM/VH state
takes about 5 days, while the full transition from LH branch to HS
branch takes more than $\sim$30 days (see Table \ref{tab:transitions}).
These properties of the BS transition can be particularly clearly seen
in XTE J1550--564 and XTE J1859+226.

GRS 1739--278 and H 1743--322 have limited statistics, but show many
resemblances to the XTE J1550--564 H-L track, transition luminosity and
duration of the transition. Alas, the distance to H 1743--322 is not
known so its luminosity was calculated for the assumed distance of 8.5
kpc and we cannot make any claims about its actual luminosity. Their HR
evolution, shown in Fig. \ref{trans}(c) and (d), is similar to the two
transitions shown in panels (a) and (b). Though errors on HR are
substantially larger in GRS 1739--278 (Fig. \ref{trans}c), its
behaviour is entirely consistent with the BS transition. H 1743--322
shows a complex HR curve with the hard-to-soft transition repeated
(Fig. \ref{trans}d) immediately after reaching the soft branch. The
first transition lasts 25 days, the total transition $\ga$80 days.

The light curves of these outbursts can be complex in shape and
generally long, extending over 200 days (with the exception of XTE
J1859+226).

\subsection{Dark/fast transition}
\label{sec:DF}

Figs. \ref{diags}(e--g) and \ref{trans}(e--g) show three outbursts with
quite different behaviour. They do not follow the XTE J1550--564 H-L
track but make a `shortcut' between the LH and HS branches at a much
lower luminosity of $\la$0.05 $L_{\rm Edd}$. The transition time is
much shorter, taking less then 16 days (Table \ref{tab:transitions}).
While the BS transients slow down their rate of transition after
reaching the IM/VH state at HR $\sim$ 1.3, the DF transients proceed to
the soft branch at the same rate (see Fig. \ref{curves}). The entire
outburst seems to be shorter then in BS state transients.

XTE J1650--500 makes the most clear example of this transition, moving
between the LH and HS branches at the luminosity of only about 0.02
$L_{\rm Edd}$ in about 16 days. The 2000 outburst of XTE J1550--564 was
very different from its 1998 outburst. The transition was much shorter
(about 7 days) and occurred at much lower luminosity. However, while
during the 1998 outburst the source reached HR = 0.4 in the soft state
(marked by a dashed line in Fig. \ref{trans}e), the hardness ratio
decreased only to 0.6 in 2000. Either it has not reached the `proper'
soft state at all, or the soft state was slightly different then in the
1998 outburst.

XTE J2012+381 is more problematic, as we do not know its distance, so
the luminosity is calculated for the assumed distance of 8.5 kpc.
However, its L-H evolution is very similar to the two other DF
outbursts, both in track shape and speed of movement. Therefore, we
classify this source as a DF transient.

Figs. \ref{diags}(h) and \ref{trans}(h) show the outburst of 4U
1543--47. It made a fast transition from the LH state to the unusually
bright HS state, which might be due to its low inclination angle (see
discussion in Sec. \ref{sec:caveats}). However, the luminosity,
duration of the transition and the evolution of hardness ratio allow us
to classify this outburst as a DF transition.

\begin{figure*}
\begin{center}
\leavevmode \epsfxsize=17cm \epsfbox{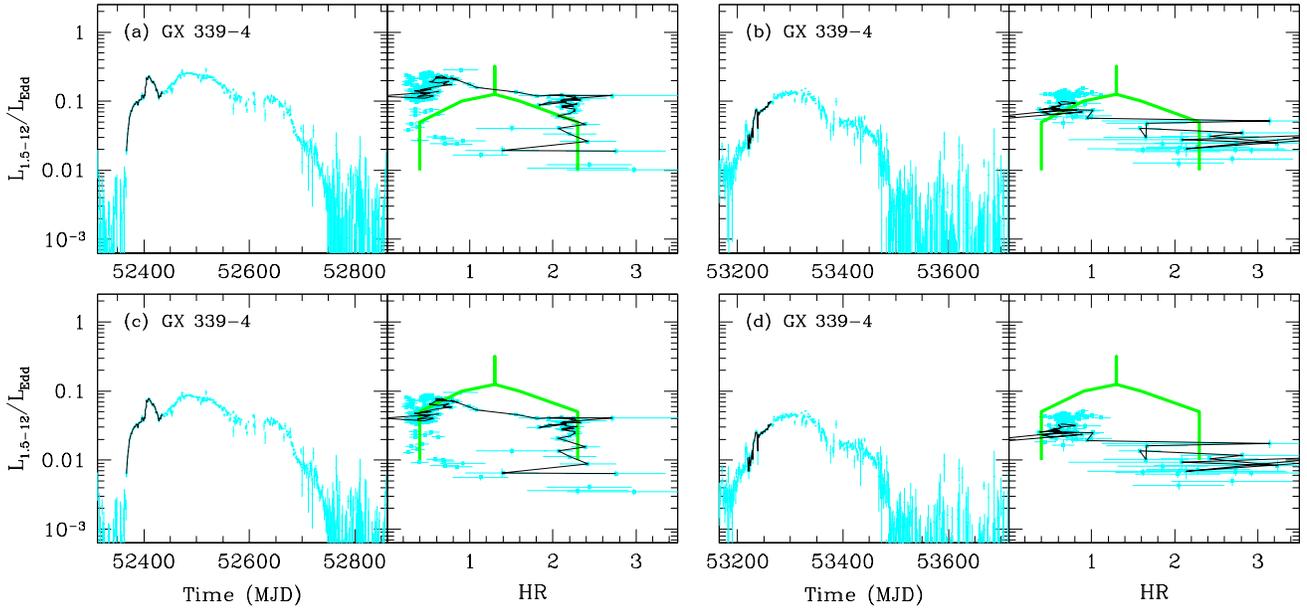}
\end{center}
\caption{Light curves and H-L diagrams of GX 339--4. Left-hand panels,
(a) and (b), show two outburst assuming the mass and distance from
Table \ref{tab:sources}, i.e. 6 M$_\odot$ and 8 kpc, respectively.
Right hand panels, (c) and (d), show the same outbursts calculated for
the mass of 10 M$_\odot$ and distance of 6 kpc. The symbols and lines
are the same as in Fig. \ref{diags}.} \label{gx}
\end{figure*}

Fig. \ref{gx} shows GX 339--4. In panel (a) we show the 2002/2003
outburst based on the most recent distance estimate of 8 kpc (Zdziarski
et al. 2004). The light curve is long and complex, resembling BS
transients. On the other hand, the shape of the H-L diagram, the
hardness ratio evolution (Fig. \ref{trans}i) and the fast transition
time (9 days) strongly suggest the DF state transition. However, the
transition luminosity $>$0.1 $L_{\rm Edd}$ is inconsistent with DF
transitions. Additionally, both LH and HS branches are unusually
bright, not consistent with any other source in our sample, except
perhaps 4U 1543--47 (Fig. \ref{diags}h). Another (2004/2005) GX 339--4
outburst, shown in Fig. \ref{gx}(b) is less luminous, and perhaps
consistent with DF transient behaviour.

In order to explain the unusual 2002/2003 outburst of GX339--4 we
consider the possibility whether it can be still consistent with other
DF transients within uncertainties of its mass and distance. Hynes et
al. (2004) gave a fairly robust lower limit on the kinematic distance
of 6 kpc. Cowley et al. (2002) gave the upper limit on the mass of 10
M$_\odot$. We use these two extremes to create a new plot, shown in
Fig. \ref{gx}(c) and (d). Clearly, both outbursts now seem to be
consistent both in the track shape and transition luminosity with other
DF transients. Though the entire light curve is more similar to other
BS transients, the {\em transition itself\/} is short and marginally
consistent with typical DF luminosity, so we tentatively classify GX
339--4 as a DF transient.

\begin{figure*}
\begin{center}
\leavevmode \epsfxsize=17.5cm \epsfbox{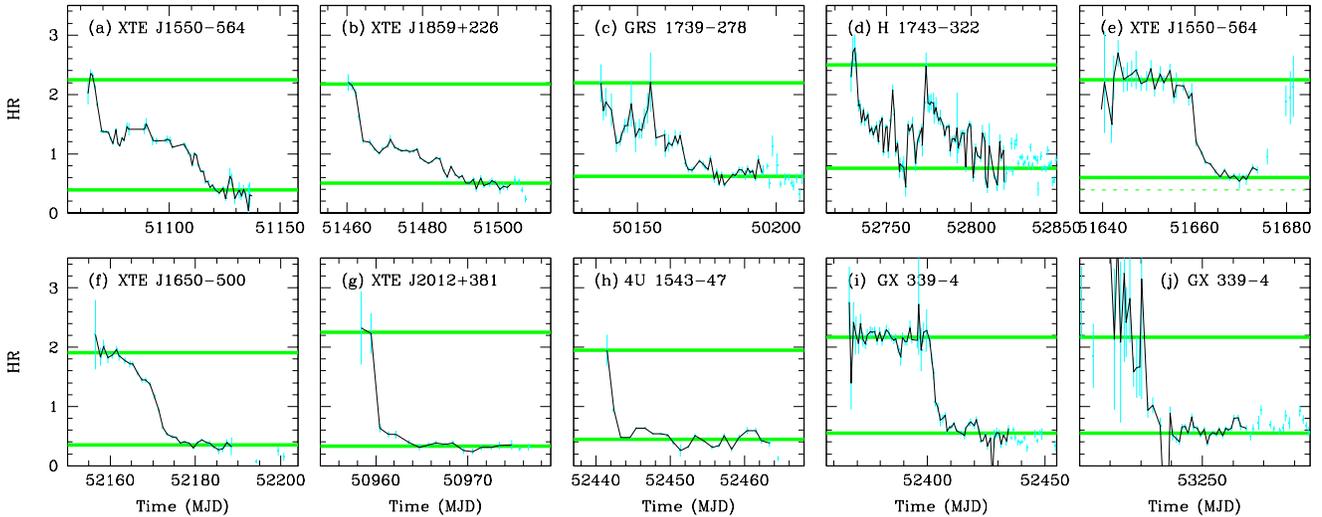}
\end{center}
\caption{Evolution of the hardness ratio during the initial part of the
outburst for several sources. The black lines connect the same points
as in Figs. \ref{diags} and \ref{gx}. The thick (green in colour)
horizontal lines show our definition of the hard (upper line) and the
soft (lower line) state for each source. The softest state of XTE
J1550--564 achieved during the 2000 outburst  (panel e) was harder then
soft state in 1998 outburst (panel a), which is marked in panel (e)
with a dashed line.} \label{trans}
\end{figure*}

\begin{table*}
\begin{tabular}{lcccc}
\hline Source Name & $L_{\rm trans}/L_{\rm Edd}$ & $t_{\rm trans}$ & Fluence &  Type\\\hline
XTE J1550--564$^{\ref{diags}a}$ & 0.13 (0.04--0.28)   & 54       & 26  & BS \\
XTE J1859+226                 & 0.12 (0.05--0.14)   & 31       & 5.7 & BS \\
XTE J1739--278                & 0.13 (0.06--0.21)   & 43       & 13  & BS \\
H 1743--322                   & [$\sim$0.17]        & 25 or $\ga$80  & [17]& BS \\
XTE J1550--564$^{\ref{diags}e}$  & 0.03 (0.01--0.07)   & 7        & 1.2 & DF \\
XTE J1650--500                & 0.02 (0.01--0.05)   & 16       & 1.3 & DF \\
XTE J2012+381                 & [0.02]              & 4        & [3.7]& DF \\
4U 1543--47                   & 0.07 (0.06--0.08)      &  2       & 9 & DF \\
GX 339--4$^{\ref{gx}a,\ref{gx}c*}$      & 0.15 (0.11--0.21) or 0.05$^*$    & 9        & 49 or 16* & DF? \\
GX 339--4$^{\ref{gx}b,\ref{gx}d*}$      & 0.06 (0.04--0.08) or 0.02$^*$    & $\sim$8   & 21 or 7*  & DF? \\
\hline
\end{tabular}

\caption{Transition luminosity (in Eddington luminosity), transition
duration (in days) and outburst fluence (in days) for sources where
transition was observed. The transition luminosity was read from the
colour-luminosity diagram at HR = 1.25, corresponding to a `cusp' in
the XTE J1550--554 diagram. The luminosity range was given for the
distance range in Table \ref{tab:sources}. The numbers in square
brackets correspond to the assumed distance of 8.5 kpc. Superscripts in
the first column denote different outbursts, referring to figure
numbers. Asterisks mark the distance to GX339--4 of 6 kpc and its mass
of 10 M$_\odot$, as opposed to the quantities listed in Table
\ref{tab:sources}}

\label{tab:transitions}
\end{table*}

\section{Caveats}
\label{sec:caveats}

In Sec. \ref{sec:results} we postulated existence of two distinct
hard-to-soft spectral transitions in the initial part of the outburst
of black hole transients. Our classification is based on the transition
luminosity, duration, the shape of the track in the H-L diagram and the
evolution of the hardness-ratio. Plainly, there are several
uncertainties here. First of all, the luminosity strongly depends on
highly uncertain distance and (to less extent) black hole mass. In
Table \ref{tab:transitions} we show the range of luminosities due to
the distance uncertainty. The uncertainties are large indeed, and there
is an overlap between the two categories around $\sim$0.05 $L_{\rm
Edd}$.

Another uncertainty comes from the interstellar absorption, affecting
the derived luminosity and hardness ratio. Fortunately, most of sources
analysed in this paper have moderate absorption column of $N_H
\la10^{22}$ cm$^{-2}$. XTE J1739--278 and H 1743--322 have higher
columns of about $2\times10^{22}$ cm$^{-2}$ (Greiner, Dennerl \&
Predehl 1996; Capitanio et al. 2005) which, depending on the shape of
the spectrum can diminish the 1.5--12 keV flux down to $\sim$0.7 of the
unabsorbed flux. The effect of this column on 3-12 keV hardness is
negligible.

The luminosities quoted in this paper are calculated in the 1.5--12 keV
energy band, but the bolometric luminosity is definitely larger. To
estimate the bolometric correction in the IM/VH state we used the same
physical model of XTE J1550--564 as in Sec. \ref{sec:data}. It gives
the bolometric luminosity 2.3 times larger then the 1.5--12 keV
luminosity. Since the IM/VH spectra are similar in shape, this seems to
be a good estimate of the bolometric correction factor. Thus, we
estimate that the bolometric luminosity is $\sim$30 per cent of $L_{\rm
Edd}$ during the BS transition and $\la$10 per cent of $L_{\rm Edd}$
during the DF transition (though 4U 1543--47 transition is brighter,
$\sim$16 per cent of $L_{\rm Edd}$, which might be due to its lower
inclination angle, see next paragraph). We have performed similar tests
for hard-state (observation id. 30188-06-01-00) and soft-state
(30191-01-43-00) data and found bolometric corrections of 2.8 and 2.5,
respectively.

The inclination angle of the accretion disc can play a significant role
in the disc-dominated HS state. The emission from optically thin
Comptonization, which dominates the LH state, is probably more
isotropic, though it might be mildly boosted at low angles if it
originates from a jet/outflow. Inclinations of the sources in our
sample range (where known) between $21^\circ$ for 4U 1543--47 (Park et
al. 2004) and $72^\circ$ for XTE J1550--564 (Orosz et al. 2002). We
expect the apparent disc in the former source to be brighter by factor
$\cos 21^\circ / \cos 72^\circ \approx 3$ then the latter one. This
might explain why the HS branch of 4U 1543--47 is about 3 times
brighter than in XTE J1550--564. The relatively low inclination of
$(40\pm20)^\circ$ (Cowley et al. 2002) cannot account for the unusually
high luminosity of GX 339--4, as it is also very bright on the LH
branch.

Opportunely, all the caveats regarding distance, absorption and
inclination angle can be dropped in the case of two different outbursts
observed from {\em the same} source, XTE J1550--564. It displayed the
BS transition in 1998 (Fig. \ref{diags}a) and the DF transition in 2000
(Fig. \ref{diags}e). Each outburst showed typical characteristics
derived for each transition category, which significantly strengthens
our result.

\begin{figure}
\begin{center}
\leavevmode \epsfxsize=8cm \epsfbox{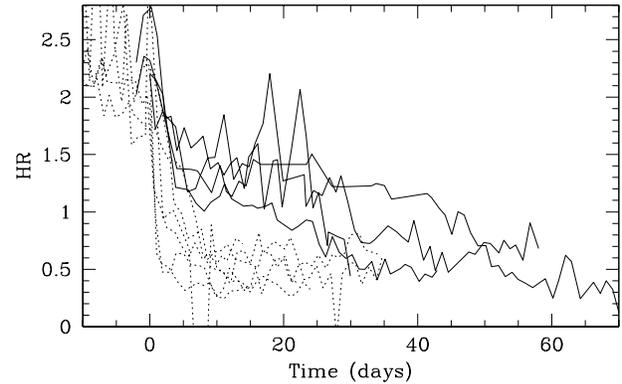}
\end{center}
\caption{Comparison of the hardness ratio evolution of all transitions
shown in Fig. \ref{trans} (the second transition of H 1743--322, after
MJD 52762, was omitted for clarity), zeroed at the beginning of the
hard-to-soft transition. Solid curves represent BS transitions and
dotted curves represent DF transitions. This figure illustrates the
crucial difference between the two types of transitions. After the
initial fast phase, the BS transition is much slower after reaching HR
$\sim$ 1.3, while DF transients continue to the soft branch at the same
rate.} \label{curves}
\end{figure}

One possible uncertainty comes from the identification of the soft- and
hard-state branches, which can subsequently affect measurements of the
transition duration. Usually, the spectral state of an X-ray source is
established on the basis of its spectral {\em and} variability
properties. In this work we define the hard and soft state branches on
the sole basis of their hardness ratio, which might not necessary agree
with the full spectral/timing definition. In particular, ASM data show
that XTE J1650-500 reached the HS branch at HR = 0.35 around MJD 52176.
We have calculated the same hardness ratio from available PCA data and
found it to be in perfect agreement with the ASM data. However, after
the transition, around MJD 52190 (see Fig. \ref{curves}f), where the
ASM data are sparse and of poor statistics, PCA shows a further
decrease in the hardness ratio to about HR = 0.15. Rossi et al. (2004)
classified the state at HR = 0.35 as the very high state, but they also
noted that its properties were quite different from the IM/VH state
during the transition. A similar situation is in the case of the 2000
outburst of XTE J1550--564, which never reached the soft state similar
to that in 1998. It is then possible, that our HS branch derived from
the ASM data contains softer IM/VH states, as defined from the spectral
and variability properties. Thus, the transition time from the hard
state to the {\em true} soft state can be longer then the transition
between the ASM LH and HS branches, derived in this paper.

This, however, does not seem to affect our main result about the
difference between the BS and DF transitions. The transition between
the LH and HS branches, {\em as derived from ASM data}, is markedly
faster and less luminous in DF transients. Even if our HS branch does
not exactly correspond to the true soft state, there is still a {\em
systematic} difference between those transitions, as illustrated in
Fig. \ref{curves}.

We would like to stress, that despite all the caveats discussed in this
section, there are two criteria distinguishing between the DF and BS
transitions, which are fairly robust. These are the duration of
transition between LH and HS branches (or evolution of the
hardness-ratio) and the shape of the L-H track. Though in some cases
the limited statistics makes recognizing the track rather difficult, we
find that when all the clues are taken together, there is a
significant, systematic difference between the two categories of
transitions. Therefore, we find our classification robust and well
defined.

\section{Discussion and conclusions}
\label{sec:discussion}

\begin{figure}
\begin{center}
\leavevmode \epsfxsize=8.5cm \epsfbox{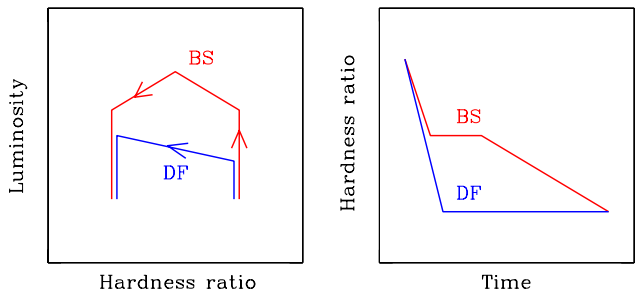}
\end{center}
\caption{Schematic diagram of two types (bright/slow and dark/fast) of
hard-to-soft transitions. Left panel: hardness-luminosity diagram.
Right panel: evolution of the hardness ratio.} \label{sketch}
\end{figure}

We have studied light curves and hardness-luminosity diagrams of
several GBH transients. We have found two distinct categories of
spectral transitions occurring after the initial hard state during the
rise of the outburst and before the disc-dominated high/soft state. The
bright/slow transition reaches the estimated bolometric luminosity of
0.3 $L_{\rm Edd}$ or more. The initial transition from the LH state is
fast, but it is significantly slowed down after reaching the bright
IM/VH state. The entire transition to the HS state takes over 30 days.
The LH and HS branches in the H-L track are connected by a roughly
semi-circular, counter-clockwise track. The dark/fast transition
typically takes place at luminosities $\la$0.1 $L_{\rm Edd}$, it is
faster (less then $\sim$15 days), does not slow down at the IM/VH state
and makes a `shortcut' between the LH and HS branches in the H-L
diagram. The difference between these two categories is shown in Fig.
\ref{sketch}.

The distinction between the two categories seems to be clear, and can
be particularly well seen in Fig. \ref{curves}. However, we only have
four transitions in each category, so from purely statistical point of
view the bimodal behaviour is not yet significant. Plainly, more data
is necessary to firmly confirm our discovery.

The terms `very high' and `intermediate' state have been used in the
literature (e.g. Miyamoto et al. 1991; Homan et al. 2001), and
sometimes the distinction between the higher and lower luminosities had
been made (e.g. M{\'e}ndez \& van der Klis 1997). Later, this
distinction was abandoned, when it became clear that the IM/VH states
can occur at various luminosities, often less than that in the
disc-dominated HS state. Instead, the definition based on both spectral
and timing properties was proposed (e.g. McClintock \& Remillard in
press). The energy and power spectra of these states are similar,
regardless of luminosity (e.g. Homan et al. 2001; Kalemci et al. 2003;
Gierli{\'n}ski \& Done 2003; Montanari, Frontera \& Amati 2004). No
systematic distinction between bright and dark IM/VH spectral states
has been found so far. In this paper we have shown that the
hard-to-soft state transition can take place at two different
luminosities, and that there is a systematic difference between the
low- and high-luminosity transitions. If physical properties of the
accretion flow are different between the DF and BS transitions, then we
might expect some difference in the spectral state as well. If this
were true than perhaps the distinction between the IM and VH states
might be valid, after all. Confirming this potential difference would
require additional studies.

There are certain limitations to the BS/DF distinction we have proposed
in this paper. The estimated BS transition bolometric luminosity of
$\sim$0.3 $L_{\rm Edd}$, observed in the four sources shown in Fig.
\ref{diags}(a--d), is not entirely typical, and can be easily
surpassed. Some of the transients reached and exceeded the Eddington
luminosity during their outbursts, e.g. V404 Cyg (e.g. {\.Z}ycki, Done
\& Smith 1999), Nova Mus ({\.Z}ycki et al. 1999), or V4641 Sgr
(Revnivtsev et al. 2002). GRS 1915+105 is also known to be very bright
(Done, Wardzi{\'n}ski \& Gierli{\'n}ski 2004), but it is a rather
untypical transient (Truss \& Done 2006). It is interesting to notice
that GX 339--4 made two transitions at significantly different
luminosity, but with time-scales and hardness ratio evolution typical
for DF transitions. Thus, the DF transition luminosity is not constant
and the transition can occur at a range of luminosities below a certain
critical value. The opposite, i.e. a range of luminosities above a
critical limit, could be true for the BS transitions. It is difficult
to ascertain using currently available data, whether these two critical
luminosities are equal or not.

We have seen both types of transitions from the same source, XTE
J1550--564. This obviously means that the outburst type is not
predestined by the properties of the binary, such as the disc size, the
orbital period or mass/spin of the black hole. The distinguishing
factor must be hidden in the accretion flow itself. The numerical model
of the disc instability by Dubus, Hameury \& Lasota (2001) shows that
two types of outbursts can develop in the accretion disc, depending on
the radius at which the hydrogen ionization instability is triggered.
When the ignition radius is small (the `inside-out' outburst) the
propagation time of the inward heating front is short, while the
outward front can easily stall in the regions of higher density, which
leads to short, low-amplitude outbursts. When the ignition radius is
large, (the `outside-in') outburst, the inward front progresses through
regions of decreasing density and can heat the whole disc, leading to
longer, brighter outbursts.

This, in principle, could explain the difference between bright and
dark transitions/outbursts. However, we can see from the light curves
that the length of the outburst is not simply correlated with its
amplitude. Though generally BS outbursts have longer and more complex
light curves then DF outbursts, there are a few exceptions. For example
XTE J1859+226 (Fig. \ref{diags}b) and XTE J2012+381 (Fig.
\ref{diags}g), have light curves very similar in shape and duration,
despite undergoing different types of LH-HS transition. GX 339--4,
tentatively categorized as a DF transient, displays very long and
complex light curves.

Perhaps a better quantity that could distinguish between `inside-out'
and `outside-in' outbursts is the fluence, which is a rough estimate of
the total amount of the accreted material. We have estimated the
fluence of each outburst (Table \ref{tab:transitions}). BS outbursts
tend to have larger fluence, though the distinction is not unique. For
example, the BS outburst of XTE J1748--288 has smaller fluence then the
DF outburst of XTE J2012+381, though distances and BH masses of both
sources are not known, so this estimate is very uncertain. And again,
GX 339--4 does not easily fit into the picture with its large fluence,
even assuming extreme black hole mass and distance, as in Sec.
\ref{sec:caveats}. Clearly, the radius at which the initial instability
occurs can be an important factor, but does not entirely explain the
observed properties of spectral transitions.

Thus, it seems that the total amount of accreted material in an
outburst does not predestine the type of the LH-HS transition. The clue
to the distinction between the DF and BS transitions might lay in the
irradiation of the disc by the central source. Irradiation can lengthen
the bright state of the disc (King \& Ritter 1998). At the same time,
it can increase evaporation rate (R{\'o}{\.z}a{\'n}ska \& Czerny 2000;
Meyer-Hofmeister, Liu \& Meyer 2005; Dullemond \& Spruit 2005), which
in turn can help sustaining the bright corona providing the source of
irradiation. An interesting feature of the observed outbursts is that
the initial transition to the IM/VH state takes roughly the same time,
irrespectively of the luminosity, and only after that transitions
diversify into slow and fast, for high and low luminosities,
respectively. We speculate, that the luminosity in the IM/VH state
defines the type of the LH-HS transition. When it reaches $\ga$0.3
$L_{\rm Edd}$ it triggers the self-sustaining bright corona, and
prolongs the bright BS transition. The dimmer and much shorter DF
transition does not reach the luminosity required to lengthen the
bright state and sustain the corona, so it disappears and the source
quickly moves on to the disc-dominated HS state.

\section*{Acknowledgements}

We thank the anonymous referee for their helpful comments and Chris
Done for stimulating discussions. MG acknowledges support through a
PPARC PDRF.


\label{lastpage}

\end{document}